\documentclass[11pt,a4paper,twoside]{article}
\usepackage[T1]{fontenc}
\usepackage[latin9]{inputenc}
\usepackage{xspace}
\usepackage{amsmath}
\usepackage{amssymb}
\usepackage{epsfig}
\usepackage{graphicx}
\usepackage{psfrag}

\usepackage[left=2cm,right=2cm,top=2.5cm,bottom=2.5cm]{geometry}

\usepackage{amsmath,amssymb,epsfig} 
\usepackage{amsfonts} 
 
\usepackage{graphicx} 
\usepackage{bm} 
\usepackage{psfrag}

\def\beq{\begin{equation}} 
\def\eeq{\end{equation}} 
\def\bed{\begin{description}} 
\def\eed{\end{description}}

\def\bea{\begin{eqnarray}} 
\def\eea{\end{eqnarray}}

\def\ba{\begin{array}} 
\def\ea{\end{array}}

\def\lsim{\mathrel{\lower4pt\hbox{$\sim$}} 
\hskip-9.5pt\raise1.6pt\hbox{$<$}\;} 
 
\def\gsim{\mathrel{\lower4pt\hbox{$\sim$}} 
\hskip-12.5pt\raise1.6pt\hbox{$>$}\;}

\begin{document}

\vspace{1.cm}

\flushright{ULB-TH/06-27}
\vspace{1.cm}

\begin{center}

{\Large {\bf  The Inert Doublet Model: an Archetype for Dark Matter}}
\vspace{0.8cm}\\
{\large  Laura Lopez Honorez, Emmanuel Nezri, Josep F. Oliver, Michel H.G. Tytgat}
\vspace{0.35cm}\\
{\it 
 Service de Physique Th\'eorique, Universit\'e Libre de Bruxelles,\\
 CP225, Bld du Triomphe, 1050 Brussels, Belgium}
\vspace{0.5cm}\\

\end{center}


\abstract{
The Inert Doublet Model (IDM), a two Higgs extension of the Standard Model with an unbroken $Z_2$ symmetry, is a simple 
and yet rich model of dark matter. We present a systematic analysis of the dark matter abundance and 
investigate the potentialities for direct and gamma indirect detection. We show that the model 
should be within the range of future experiments, 
like GLAST and ZEPLIN. The lightest stable 
scalar in the IDM is a perfect example, or archetype of a weakly interacting massive particle. 
}

\section{Introduction} 
 
Contemporary cosmological observations concur to indicate that 
the majority of matter in the universe not only does not shine but is not even made of ordinary atoms \cite{Spergel:2006hy,Seljak:2006bg}. 
Deciphering the nature of this so-called Dark Matter 
has become one of the most important issue at the frontier 
of particle physics, astrophysics and cosmology. 
A profusion of dark particles have been 
proposed over the years and it is 
much hoped that present and forthcoming experiments will 
throw some light 
on the matter. For a review, see for instance \cite{Bertone:2004pz,Jungman:1995df}.
 
In the present article we investigate further a rather 
mundane, although in our opinion quite interesting, form of dark matter. The particle 
candidate is a weakly interacting massive scalar and it has been 
advocated recently in \cite{Ma:2006km} and \cite{Barbieri:2006dq,Cirelli:2005uq}. The 
framework is that of a two Higgs doublets, $H_1$ and $H_2$, version of the Standard Model with a $Z_2$ 
symmetry such that 
$$ 
H_1 \rightarrow H_1 \;\; \mbox{\rm and}\;\; 
H_2\rightarrow - H_2. 
$$ 
All the fields of the Standard Model are even under $Z_2$. This model has been first discussed by Deshpande and Ma in \cite{Deshpande:1977rw}. Following \cite{Barbieri:2006dq}, we will assume that $Z_2$ is not spontaneously broken, {\em i.e.} $H_2$ does 
not develop an expectation value. Among other things, the discrete 
symmetry prevents the appearance 
of flavour changing neutral currents. The model is definitely not the $\langle H_2\rangle \rightarrow 0$ limit 
of a generic two Higgs model like, for instance, the Higgs sector of the MSSM. 
 
The most general, albeit renormalizable, potential of the model 
can be written as 
\begin{equation} 
\label{potential} 
V = \mu_1^2 \vert H_1\vert^2 + \mu_2^2 \vert H_2\vert^2  + \lambda_1 \vert H_1\vert^4 + 
 \lambda_2 \vert H_2\vert^4 + \lambda_3 \vert H_1\vert^2 \vert H_2 \vert^2 
 + \lambda_4 \vert H_1^\dagger H_2\vert^2 + {\lambda_5\over 2} \left[(H_1^\dagger H_2)^2 + h.c.\right]. 
\end{equation} 
There is an Peccei-Quinn $U(1)$ global 
symmetry if $\lambda_5 =0$. This limit is however not favored 
by dark matter direct detection experiments (section \ref{sec:DD}).
 
The $SU(2) \times U(1)$ symmetry is broken by the vacuum expectation value of $H_1$, 
$$\langle H_1\rangle = {v\over \sqrt 2}
$$ 
with $v = -\mu_1^2/\lambda_1 = 248$ GeV 
while, assuming $\mu_2^2 > 0$, 
$$ 
\langle H_2\rangle = 0. 
$$ 
The mass of the Brout-Englert-Higgs 
particle ($h$ or Higgs for short) is 
\begin{equation}
M_h^2 = - 2 \mu_1^2 \equiv 2 \lambda_1 v^2 
\label{hmass}
\end{equation}
while the mass of the 
charged, $H^+$, and two neutral, $H_0$ and $A_0$, components of the field $H_2$ are given by 
\begin{eqnarray} 
M_{H^+}^2 &=& \mu_2^2 + \lambda_3 v^2/2\cr 
M_{H_0}^2 &=& \mu_2^2 +  (\lambda_3 + \lambda_4 + \lambda_5) v^2/2\cr 
M_{A_0}^2 &=& \mu_2^2 +  (\lambda_3 + \lambda_4 - \lambda_5) v^2/2. 
\label{masses} 
\end{eqnarray} 
For appropriate quartic couplings, $H_0$ or $A_0$ is the 
lightest component of the $H_2$ doublet.
 In the absence of any other lighter $Z_2$-odd field, 
either one is a candidate for dark matter. 
For definiteness we choose $H_0$. All our 
conclusions are unchanged 
if the dark matter candidate is $A_0$ instead. 
Following \cite{Barbieri:2006dq} we parameterize the contribution from 
symmetry breaking to the mass of $H_0$ by $\lambda_L=(\lambda_3 + \lambda_4 + 
\lambda_5)/2$, which is also the coupling constant between the Higgs field $h$
  and our dark matter candidate $H_0$. 
Of the seven parameters of the potential (\ref{potential}), one is known, $v$, 
 and four can be related 
to the mass of the scalar particles. In the sequel, we take 
$\mu_2$ and $\lambda_2$ as the last two independent parameters. 
The later actually plays little role for the question of dark matter. 
 
The abundance of $H_0$ as the dark matter of the universe as well as 
the constraints from  direct detection experiments 
have been touched upon in the recent 
literature \cite{Barbieri:2006dq,Cirelli:2005uq} (see also \cite{Majumdar:2006nt}). 
In the present article we investigate the potentialities 
 for indirect detection of $H_0$ using gamma ray telescopes and detectors. 
For the sake of completeness, we also present 
a systematic (albeit tree-level) analysis of the abundance of the $H_0$ 
 in the light of WMAP data, as well as 
 the prospect for direct detection by underground detectors.

Before doing so, we would like to 
briefly emphasize some of the virtues of this, so-called, Inert Doublet Model (IDM), both 
in general and for the issue of dark matter in particular. 
True, the model is {\em ad hoc}. Also, it does not address any 
deep issue, say the hierarchy problem. Yet it is a very simple extension of the Standard Model and 
its phenomenology is nevertheless very rich \cite{Ma:2006km,Barbieri:2006dq,Casas:2006bd,Calmet:2006hs}.
In reference \cite{Barbieri:2006dq}, Barbieri {\em et al} considered this model as a way of pushing 
the mass of the Higgs particle toward the TeV scale, without 
contradicting LEP precision measurements (see 
also \cite{Casas:2006bd}, and section \ref{sec:constr} of the present paper). 
Also, in the simplest version discussed here there are no Yukawa couplings 
to the $H_2$ doublet, but it is somewhat natural to imagine coupling 
 it to (odd) right-handed neutrinos. This opens the 
possibility of generating the mass of the SM neutrinos through loop corrections, a 
mechanism introduced by Ma in \cite{Ma:2006km} and 
further addressed in a series of papers \cite{Kubo:2006yx,Hambye:2006zn,Ma:2006fn}. 
 
As a dark matter candidate, $H_0$ is not without interest either. It belongs to the family of 
Weakly Interacting Massive Particles (WIMP) and has been proposed as a dark matter candidate by Ma . The most acclaimed member of this family 
is certainly the neutralino, the lightest supersymmetric particle (LSP). The 
supersymmetric extensions of the SM are without doubt very 
 interesting and well motivated but they 
involve many new parameters and their phenomenology is, to say the least, complex. 
It is perhaps far fetched to compare the respective advantages 
 and disadvantages of the IDM and of the MSSM but it is interesting that the 
$H_0$ shares many similarities 
 with the LSP (weak interactions, similar mass scale, etc) while being considerably 
much simpler to analyze. 
Interestingly, if we take seriously the idea that 
$H_0$ is the dominant form of dark matter, the parameter space of the IDM model 
is quite constrained. It is also, to some extent, 
complementary to that of the MSSM, as we show in the conclusions. 
And, cherry on the top, the phenomenology of the $H_0$ as a candidate for dark matter 
is neatly intertwined with that of the Higgs particle. 
In our opinion, it is this conjunction of simplicity 
and richness that makes the lightest stable scalar a perfect example, or archetype, of dark matter.

The plan of the paper is as follow. In the next section (\ref{sec:generalities}) 
we give a summary of our methodology for the calculations of the dark matter abundance, the 
photon flux 
for 
indirect detection by gamma ray telescopes and the cross-sections for direct detection. 
We then discuss in some details the results of our analysis, in the light of WMAP data, and 
study the prospects for both indirect and direct detection by existing and forthcoming experiments. 
We give our conclusions in the last section, together  with comparison of the
Model with that of the MSSM, and the prospects for future analysis.

\section{Dark matter aspects} 
 
\label{sec:generalities} 
 
\subsection{Dark Matter abundance}\label{sec:abundance} 
 
The abundance of $H_0$ has been estimated in \cite{Barbieri:2006dq} and in
\cite{Cirelli:2005uq} for two different mass ranges. 
As in these papers, we 
 consider only the standard freeze-out mechanism. 
There are then essentially two regimes, depending on the mass of $H_0$ 
 with respect to that of the $W$ and $Z$ bosons. 
If $M_{H_0} \gsim 80 {\rm GeV}$, the $H_0$ annihilate essentially into Z and W
boson pairs (see Figure (\ref{feyn:WZ})).
For $M_{H_0} > M_h$, the annihilation channel
into $h$ pairs opens (see Figure (\ref{feyn:higgs})).
Otherwise, the $H_0$ annihilate essentially 
 through an intermediate Higgs, provided the Higgs itself is not too heavy. 
Some amount of coannihilation of $H_0$ with the next-to-lightest scalar particle $A_0$ (resp. $H^+$) 
through a $Z$ (resp. $W^+$) boson can be present \cite{Griest:1990kh}, a feature that 
substantially complicates the determination of 
the dark matter abundance (see
 Figure (\ref{feyn:ffbar})).  
 
We have computed the relic 
abundance of $H_0$ using micrOMEGAs2.0, a new and versatile 
package for the numerical calculation 
of Dark Matter abundance from thermal freeze-out \cite{Belanger:2006is}. 
The latest implementation of this code, 
which was originally developed 
to study supersymmetric models, allows 
one to enter any  model containing a 
discrete symmetry that guarantees the stability of 
the dark matter particle. The code takes advantage of the fact that 
each odd ({\it i.e.} non standard) particles 
of the model will eventually decay into the lightest odd particle, in our case the $H_0$. 
One can then sum the system of Boltzmann 
equations  of all odd species and the relic density of dark matter is 
 calculated by solving the Boltzmann equation for the lightest particle only 
 \cite{Edsjo:1997bg}: 
\begin{equation} 
\frac{dY}{dT}=\sqrt{\frac{\pi g_*(T)}{45}}M_{Pl} \langle \sigma v \rangle (Y^2(T)-Y^2_{eq}(T)), 
\label{eq:boltz} 
\end{equation} 
where $Y\equiv n_{DM}/s$ is the comoving density of dark matter. 
All the processes of the model enter in the thermally 
averaged cross-section $\langle \sigma v \rangle$. 
Integrating Eq.(\ref{eq:boltz}) from $T=\infty$ to $T=T_0$ the relic abundance is given by 
\begin{equation} 
\Omega_{DM} h^2=2.72 \times 10^8 \frac{M_{DM}}{{\rm GeV}} Y(T_0). 
\end{equation} MicrOMEGAs2.0 itself is build upon CALCHEP, a code for 
computing tree-level cross-sections. We have used both MicrOMEGAs2.0 and 
CALCHEP to compare with 
our analytical calculations and to perform a number of cross-checks that have 
 strengthened our faith in the numerical results reported in the present paper.\footnote{The Feynman rules and definition files of the IDM to be used with micrOMEGAs 
 can be obtained upon request to the authors of 
the present article and will 
be included in the 
next available version of the code.}

\subsection{Indirect detection}\label{sec:indirect}

The 
measurement of secondary particles coming from dark matter 
annihilation in the halo of the galaxy is a promising way of deciphering the nature of dark matter. This 
possibility depends however not only on the properties of the dark matter particle, through its annihilation cross-sections, 
but also on the astrophysical 
assumptions made concerning the distribution of dark matter in the halo that supposedly surrounds our galaxy. 
The galactic center (GC) region is potentially a 
very attractive target for indirect detection of dark matter, in particular through gamma 
rays. The produced gamma ray flux from the annihilation of dark matter particles can be 
expressed as 
\begin{equation} 
\frac{\Phi_{\gamma}}{d \Omega d E}= \sum_i 
\frac{dN^i_{\gamma}}{dE_{\gamma}} \langle \sigma_i v \rangle \frac{1}{4 \pi m_{DM}^2} 
\int_{\mbox{l.o.s.}} 
\rho^2 \,\, d l , 
\label{flux} 
\end{equation} 
where $m_{DM}$ is the dark matter particle mass, $\rho$ is the dark matter density profile, 
$\langle \sigma_i v \rangle$ and  $dN^i_{\gamma}/dE_{\gamma}$ are, respectively, the
thermally averaged 
annihilation cross section times the relative velocity $v$ and  the differential gamma spectrum  per 
annihilation coming from the decay of annihilation products of final state $i$. 
The integral is taken along the line of sight.
The processes involved are shown in the Figure (\ref{feyn:ffbar}), first diagram and Figures (\ref{feyn:WZ}) and
(\ref{feyn:higgs}).
 
The models will be constrained by the existing EGRET \cite{EGRET} experimental limit on the flux, $\sim 
10^{-8}$ photons ${\rm cm^{-2}.s^{-1}}$, and the forthcoming GLAST \cite{GLAST} expected sensitivity, 
$\sim 10^{-10}-10^{-11}$ photons ${\rm cm^{-2}.\,s^{-1}}$ for $\Delta \Omega=10^{-3}$ and 
$10^{-5}$ srad respectively. 
 
Our purpose for the time being is merely to prospect the IDM with regard to gamma ray 
indirect detection. In this paper, we focus on the particle physics parameter 
dependence of the model and work within a fixed astrophysical framework, the 
popular Navarro-Frank-White (NFW) halo profile \cite{nfw}. 
With $R_0=8.5$ kpc equal to the distance from to Sun to the GC 
and $\rho_0=0.3{\rm \ GeV.cm^{-3}}$ the dark matter density in our 
neighborhood, the NFW profile density can be parameterized as 
\begin{equation} 
  \rho(r)= \rho_0 \frac{[1+(R_0/a)^{\alpha}]^{(\beta-\gamma)/\alpha}} 
  {[1+(r/a)^{\alpha}]^{(\beta-\gamma)/\alpha}} 
  \left(\frac{R_0}{r}\right)^{\gamma}  \;\;, 
\label{eq:alphabetagamma} 
\end{equation} 
with $(a,\alpha,\beta,\gamma)=(20,1,3,1)$. It should be emphasized that the (dark) matter 
distribution in the innermost region of the galaxy 
is poorly known so that 
$\gamma$ is not very constrained, a freedom that can give rise to very different 
values of the gamma ray flux. Typically a suppression or an enhancement of two 
orders of magnitude can be obtained if one considers respectively a halo with 
a flat core ({\it e.g.} isothermal, $\gamma=0$) or a deeper cusp 
($\gamma\sim1.5$) \cite{moore} coming for instance from baryonic infall. Hence, 
depending on the astrophysics assumptions, the estimation of gamma ray signal can vary 
quite strongly (see {\em e.g.} \cite{compression,lia}). 
 
To calculate the flux we have integrated  (\ref{flux}) for a NFW profile above $1 
{\rm GeV}$ and around a solid angle of $\Delta \Omega=10^{-3}$ srad for EGRET and 
$\Delta \Omega = 10^{-5}$ srad for GLAST. The differential spectra of each channel are given by 
micrOMEGAs (we have checked the agreement with Pythia simulations of
\cite{compression}) as well as $\langle \sigma v \rangle$ at rest and we have integrated 
the square of the dark matter density along the line of sight. 
 
\subsection{Direct detection}\label{sec:DD} 
 
A local distribution of weakly interacting dark matter could be detected \cite{Goodman:1984dc}
by measuring the energy deposited in a low background detector by the scattering of a dark
 matter particle with a nuclei of the detector. 
One distinguishes spin dependent and spin independent interactions. 
For $H_0$ interacting with the quarks of the nuclei, there are two spin
independent processes at tree level, 
$H_0 q \xrightarrow{Z}A_0 q$ and  $H_0 q \xrightarrow{h}H_0 q$ (see Figure (\ref{feyn:DD})). 
The experiments have reached such a level of sensitivity that the $Z$ exchange contribution is excluded by the current 
experimental limits. Consequently, to forbid $Z$ exchange by kinematics, the mass of the $A_0$ particle must be
higher than the mass of $H_0$ by a few $100 {\rm keV}$ (we will thus disregard  the $\lambda_5 \rightarrow 0$ limit). 
We assume that the main  process here is the one with Higgs particle exchange $h$.
The cross section at tree-level is \cite{Barbieri:2006dq} 
\begin{equation} 
\sigma^h_{H_0-p}=\frac{m_r^2}{4 \pi}\left(\frac{\lambda_L}{M_{H_0}M^2_h}\right)^2 f^2 m^2_p, 
\label{eq:sigmaDMp} 
\end{equation} 
where $f$ is a form factor estimated in the literature to be $f\sim0.3$ and $m_r$ is the reduced mass of the system. 
\cite{Ellis:2000ds}. 
In addition, for $M_{H_0}<M_W$, we have neglected the one-loop exchange of two gauge bosons. 
For $M_{H_0}>M_W$, we include
 this process with, following
\cite{Cirelli:2005uq}, 
$\sigma_{H_0-p}~\simeq~4.6\,10^{-13}{\rm pb}$. 
We take into account the constraints given by the CDMS \cite{CDMS} 
experiment ($\sim 10^{-6}$ pb) and the model is also compared 
to the next generation experiments like EDELWEISS II \cite{EdeII} or 
a ton-size experiment like ZEPLIN \cite{Zeplin} with the valley of 
their sensitivities respectively around $\sigma \sim 10^{-8}$ and  $\sigma \sim 10^{-10}$ pb. 
 
\section{Analysis} 
 
\begin{figure} 
\begin{center} 
 \begin{tabular}{ccc} 
\psfrag{H}[c][l]{$H_{0}$} 
\psfrag{W}[c][tc]{$W^+\, (Z)$} 
\psfrag{W2}[tc][bc]{$W^-\, (Z)$} 
\includegraphics[width=0.18\textwidth]{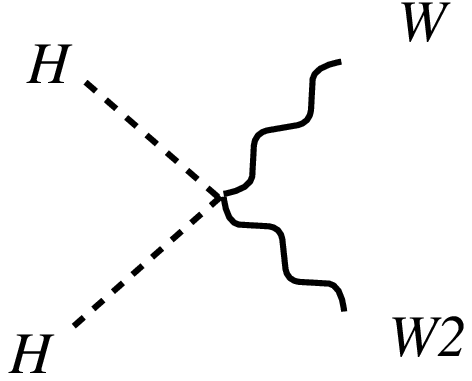}\hspace*{0.07\textwidth}& 
\psfrag{H}[c][c]{$H_{0}$} 
\psfrag{Hp}[l][c]{$H^-(A_0)$} 
\psfrag{W}[c][c]{$W^+\,(Z)$} 
\psfrag{W2}[c][b]{$W^-\,(Z)$} 
\includegraphics[width=0.2\textwidth]{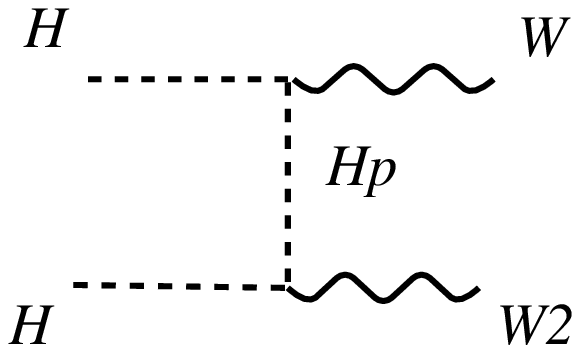} \hspace*{0.07\textwidth}& 
\psfrag{H}[c][l]{$H_{0}$} 
\psfrag{h}[c][c]{$h$} 
\psfrag{W}[c][c]{$W^+ \,(Z)$} 
\psfrag{W2}[c][c]{$W^-\, (Z)$} 
\includegraphics[width=0.2\textwidth]{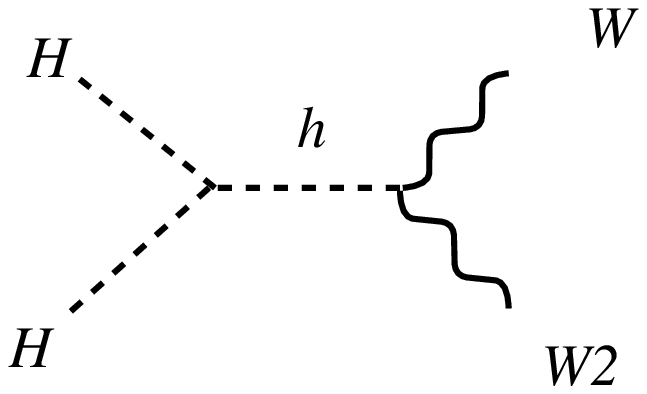}\\ 
 $g^2$ & 
 $g^2$ & 
 $\lambda_L g$ 
\end{tabular} 
 \caption{\small Annihilation channels into gauge bosons final state with corresponding couplings.} 
\label{feyn:WZ} 
   \end{center} 
\end{figure}

\begin{figure}
\begin{center} 
 \begin{tabular}{ccc} 
\psfrag{H}[c][c]{$H_{0}$} 
\psfrag{h}[c][c]{$h$} 
\includegraphics[width=0.18\textwidth]{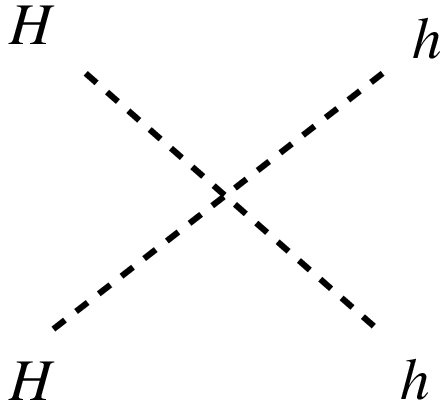}& 
\psfrag{H}[c][c]{$H_0$} 
\psfrag{h}[c][c]{$h$} 
\includegraphics[width=0.2\textwidth]{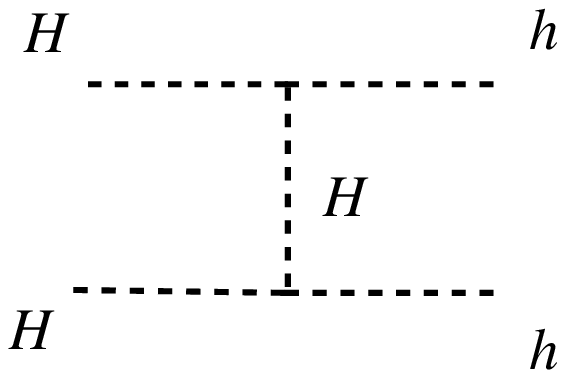}& 
\psfrag{H}[c][c]{$H_0$} 
\psfrag{h}[c][c]{$h$} 
\includegraphics[width=0.2\textwidth]{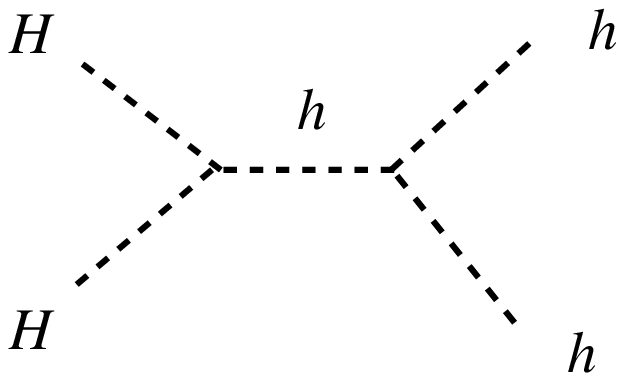}\\ 
 $\lambda_L$ & 
 $\lambda_L^2$ & 
 $\lambda_L \lambda_1$ 
\end{tabular} 
 \caption{\small Annihilation channels into Higgs final state.} 
\label{feyn:higgs} 
   \end{center} 
\end{figure}

\begin{figure}
\begin{center} 
 \begin{tabular}{lr} 
\psfrag{H}[c][c]{$H_{0}$} 
\psfrag{h}[c][c]{$h$} 
\psfrag{f}[c][c]{$f$} 
\psfrag{f2}[c][c]{$\bar{f}$} 
\includegraphics[width=.2\textwidth]{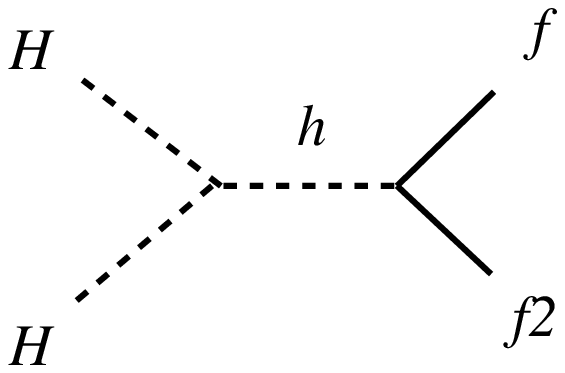}& \hspace*{0.07\textwidth} 
\psfrag{H}[c][l]{$H_{0}$} 
\psfrag{A}[rc][lc]{$A_{0} (H^+)$} 
\psfrag{Z}[c][c]{$Z (W^+)$} 
\psfrag{f}[c][c]{$f^{(')}$} 
\psfrag{f2}[c][c]{$\bar{f}$} 
\includegraphics[width=0.2\textwidth]{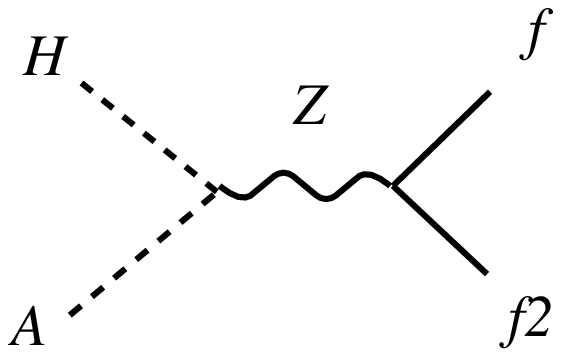}\\ 
 $\lambda_L y_f$ & 
 $g^2$ 
\end{tabular} 
 \caption{\small (Co)Annihilation channels into fermion anti-fermion final state.} 
\label{feyn:ffbar} 
   \end{center} 
\end{figure}

\begin{figure}[h!]
\begin{center} 
 \begin{tabular}{cc} 
\psfrag{H}[c][c]{$H_{0}$} 
\psfrag{h}[c][c]{$h$} 
\psfrag{f}[c][c]{$f$} 
\psfrag{f2}[c][c]{$\bar{f}$} 
\includegraphics[width=.23\textwidth]{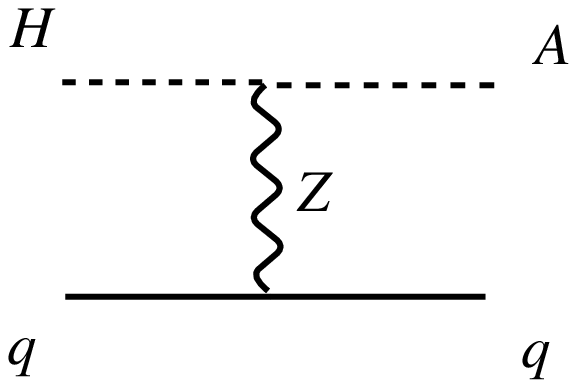}& 
\psfrag{H}[c][c]{$H_{0}$} 
\psfrag{A}[c][c]{$A_{0}$} 
\psfrag{Z}[c][c]{$Z$} 
\psfrag{f}[c][c]{$f$} 
\psfrag{f2}[c][c]{$\bar{f}$} 
\includegraphics[width=0.23\textwidth]{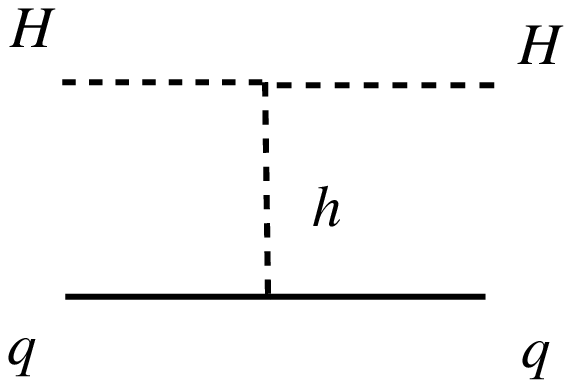}\\ 
 $\lambda_L y_f$ & 
 $g^2$ 
 
\end{tabular} 
 \caption{\small Leading channels contributing to $\sigma_{H_0-p}$ (direct detection). } 
\label{feyn:DD} 
   \end{center} 
\end{figure}

The processes driving the (co)annihilation cross section $\langle \sigma v \rangle$ relevant 
for the relic density from freeze-out and indirect detection of gamma rays 
are shown in the Figures (\ref{feyn:WZ}),  
(\ref{feyn:higgs}) and (\ref{feyn:ffbar}). The $H_0$-proton interactions relevant for direct detection are shown in 
 Figure (\ref{feyn:DD}). 
 
There are two qualitatively distinct regimes, depending on whether the $H_0$ 
is lighter than the $W$ and $Z$ and/or the Higgs boson, in which case  
annihilation proceeds through the diagrams of Figure (\ref{feyn:ffbar}). At low mass, the second diagram of (\ref{feyn:ffbar})
may contribute to the dark matter 
abundance if there is a substantial number of $A_0$ at the time of freeze-out ({\em i.e.} coannihilation). 
 
In the plots of Figure (\ref{beW}), the relic 
abundance of $H_0$ dark matter particles, the gamma ray flux due to $H_0$ annihilation at the galactic center 
  (indirect detection) 
and, finally,  their cross-section for elastic scattering off a proton through Higgs boson 
$\sigma_{H_0-p}$ (direct detection) 
exchange are shown, 
for two particular Higgs masses ($M_h = 120$ GeV and $200$ GeV) in the ($M_{H_0},\mu_2$) plane. We refer to this as the low $H_0$ mass regime. For higher $H_0$ masses, 
similar plots are displayed in Figure (\ref{abW}) (for  $M_h = 120$ GeV), respectively for the dark matter abundance, 
the gamma ray flux and the cross-section for direct detection. 
In each of these three cases, the color gradients correspond respectively to gradients in 
   $\log(\Omega_{H_0}h^2)$, $\log( 
   \Phi_\gamma (\mbox{cm}^{-2} \mbox{s}^{-1}))$ and $\log( \sigma_{H_0-p}(\mbox{pb}))$. 
 
Since we plot our results in the ($M_{H_0},\mu_2$) plane, 
the diagonal line corresponds to $\lambda_L =0$, {\em i.e. to no coupling} between $H_0$ and the Higgs boson. 
Away from this line, $\lambda_L$ increases, with $\lambda_L <0$ (resp. $\lambda_L > 0$) above (resp. below) 
the diagonal. 
Also, we write $\Delta MA_0= M_{A_0}-M_{H_0}$ and  $\Delta MH_c= M_{H^+}-M_{H_0}$. 
 
In the plots of the dark matter abundance, the areas between the two dark lines correspond to regions 
of the parameter space such that $0.094<\Omega_{DM}h^2< 0.129$, the range of 
 dark matter energy densities consistent with WMAP data. 
In the cross section and gamma ray flux plots, the lines indicate the areas of the parameter 
space within reach of the various experiments we take in consideration. 
 
\subsection{Constraints } 
 \label{sec:constr}
 
\begin{figure} 
\begin{center} 
\begin{tabular}{cc} 
\includegraphics[width=0.47\textwidth]{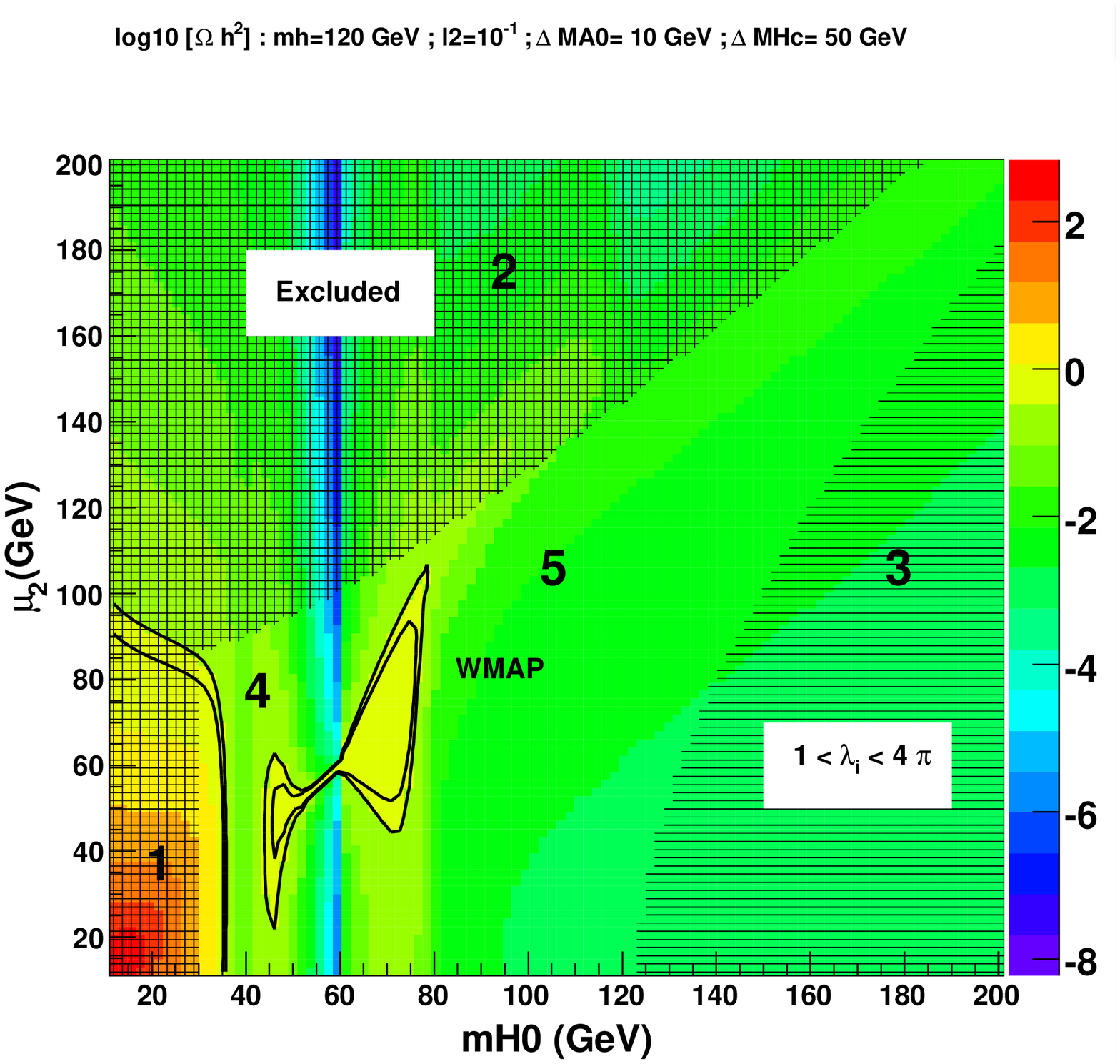}& 
\includegraphics[width=0.47\textwidth]{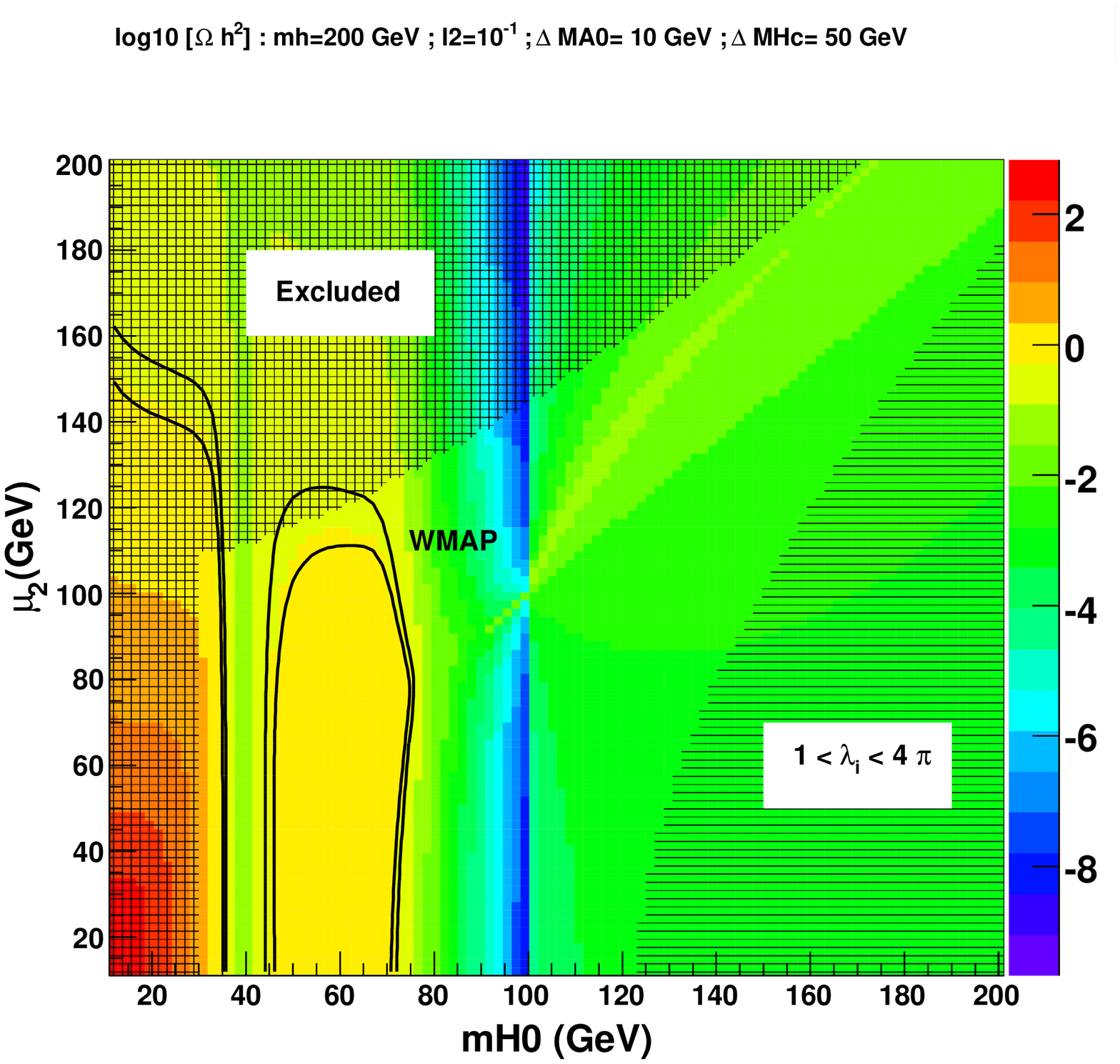}\\ 
\includegraphics[width=0.47\textwidth]{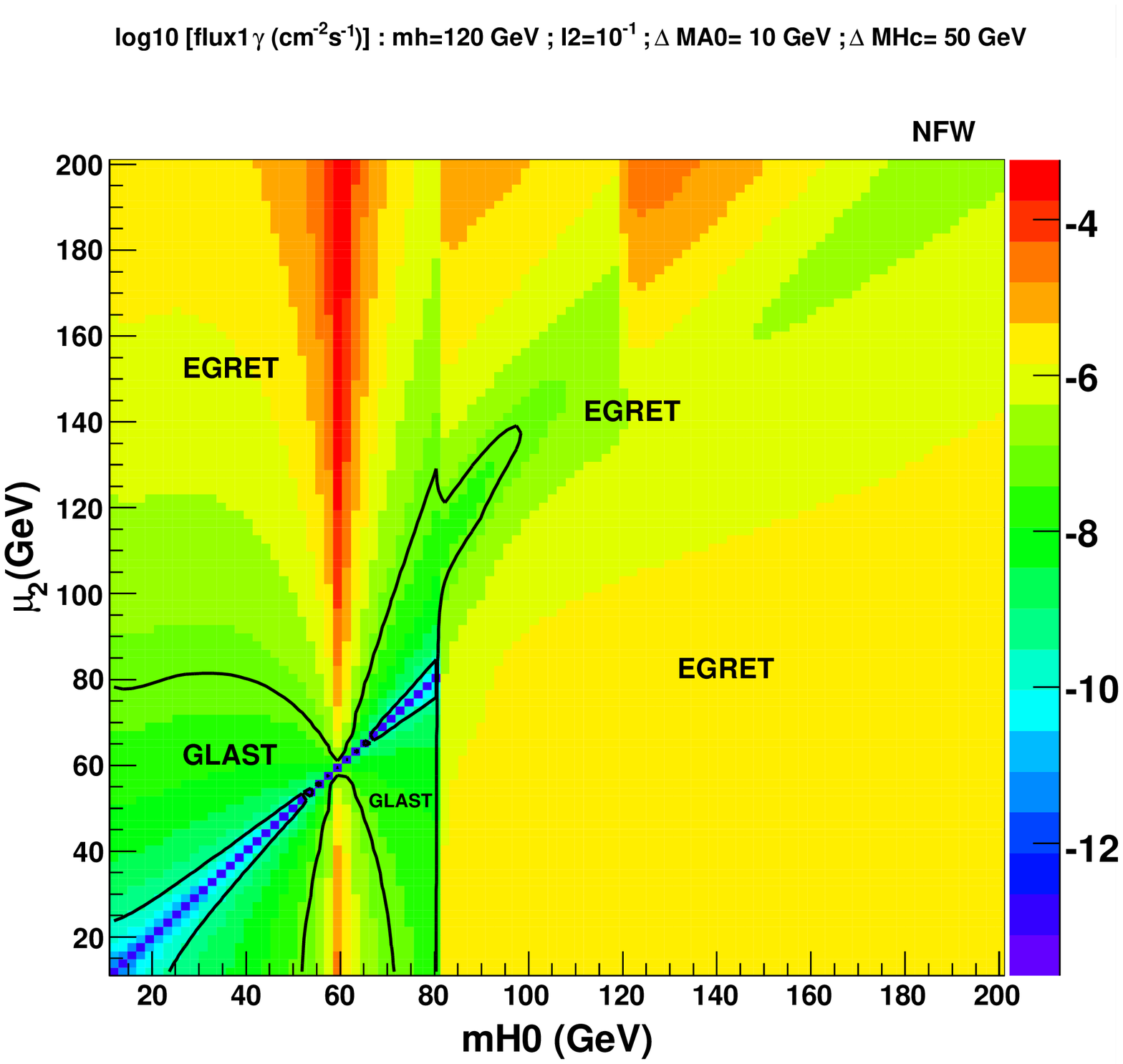}& 
\includegraphics[width=0.47\textwidth]{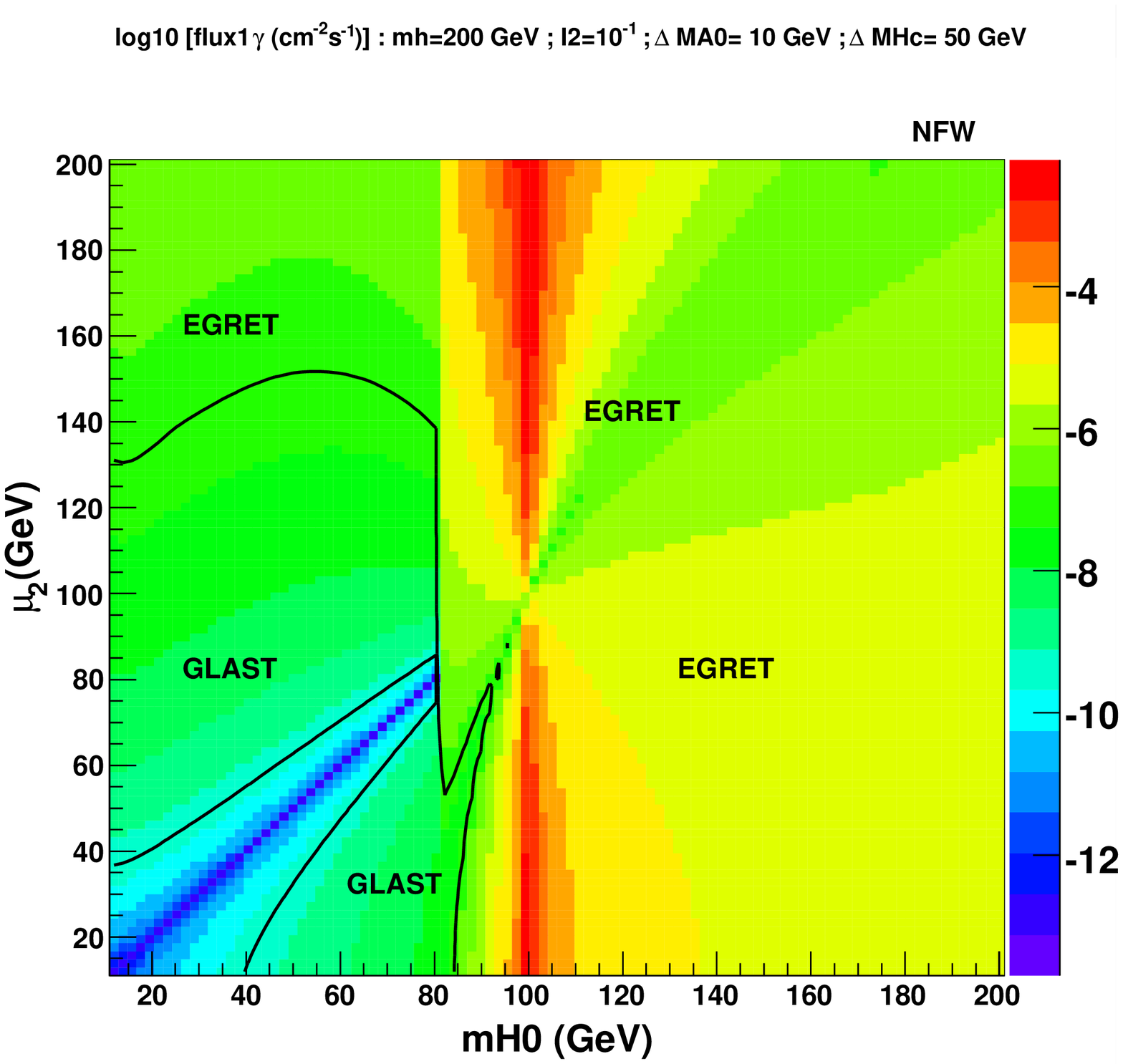}\\ 
\includegraphics[width=0.47\textwidth]{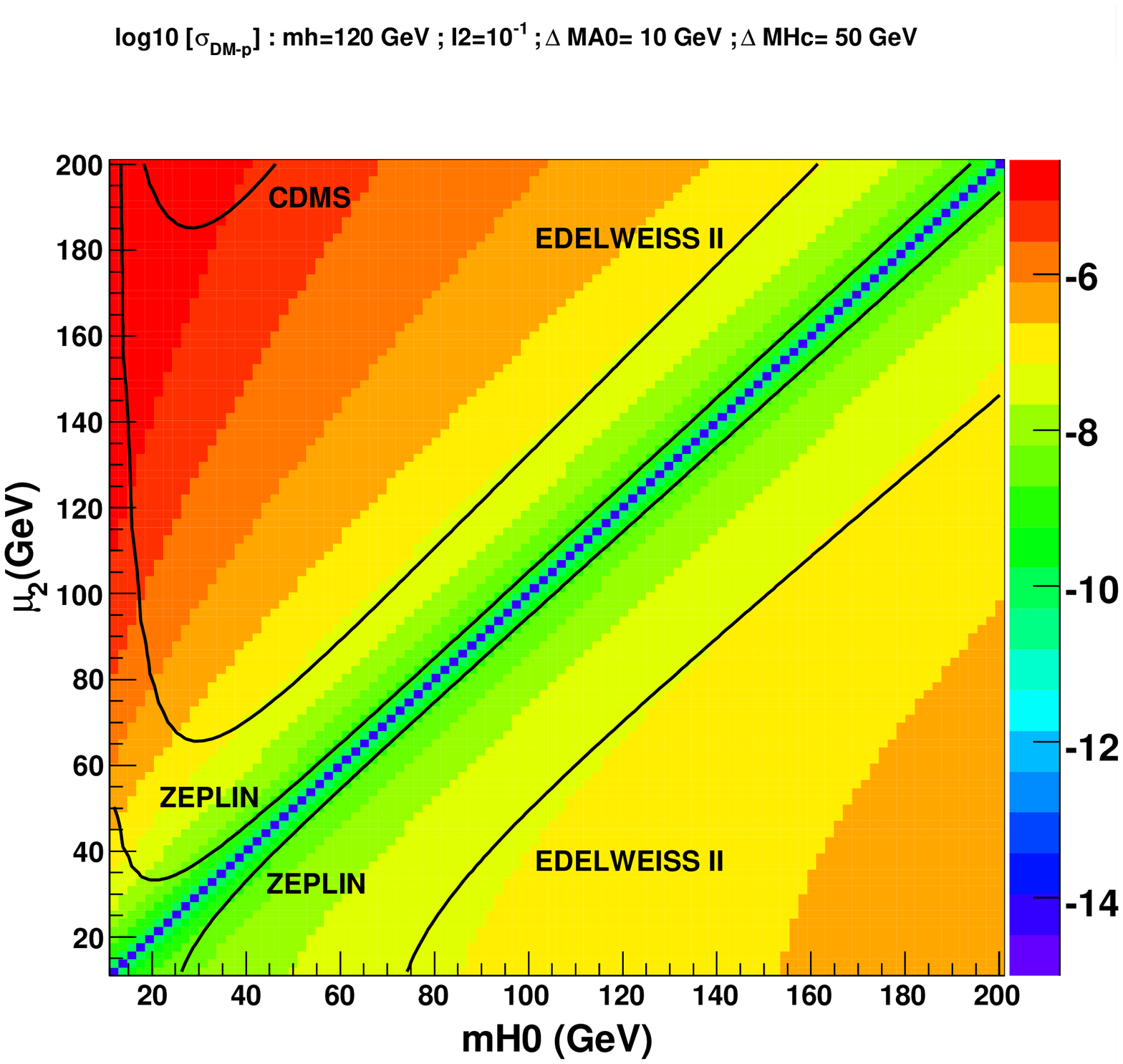}& 
\includegraphics[width=0.47\textwidth]{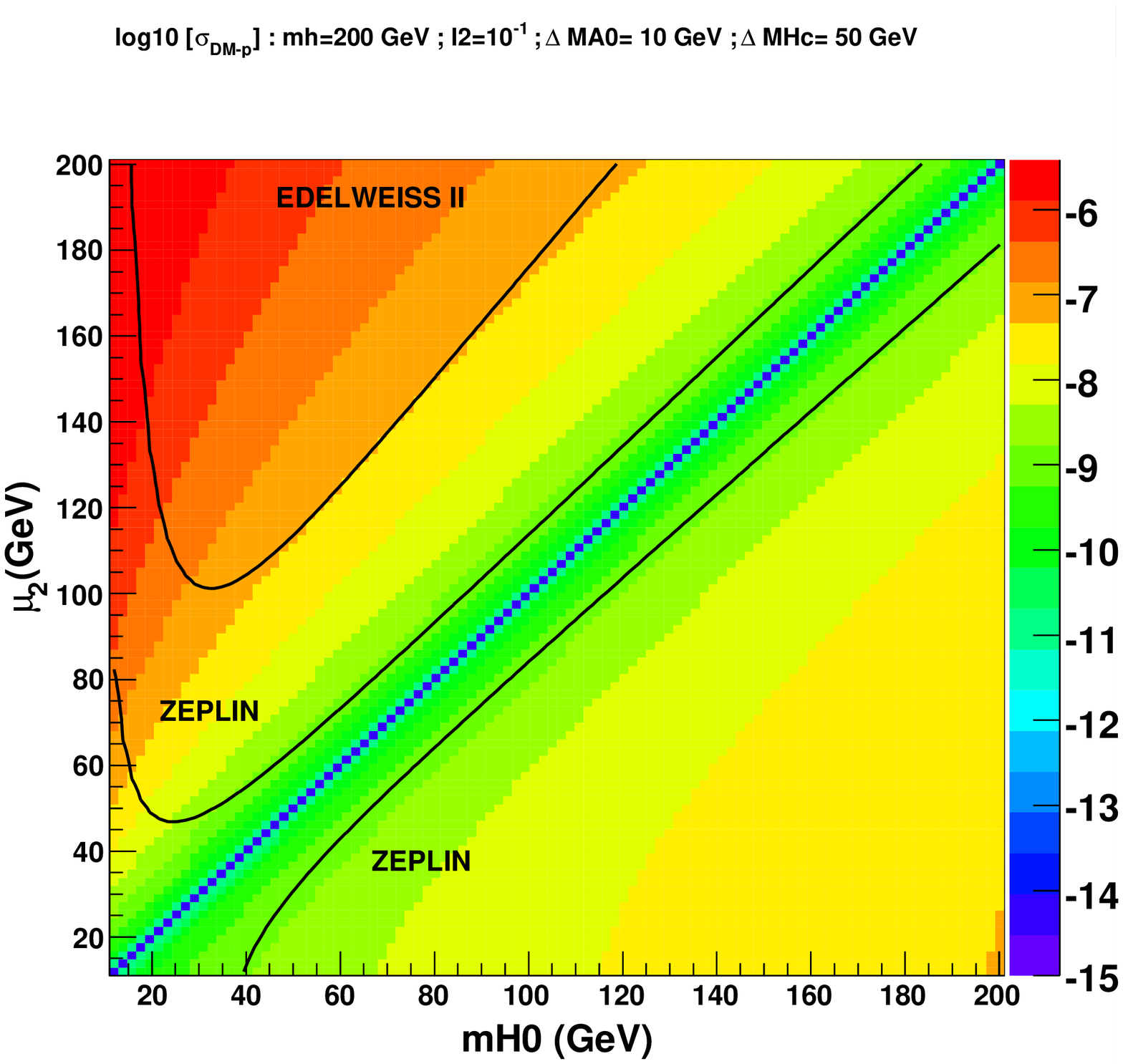} 
\end{tabular} 
\caption{From top to bottom: Relic density, gamma indirect detection and
  direct detection contours in the $(M_{H_0},\mu_2)$ plane. Left: $m_h=120$
  GeV.  Right: $m_h=200$ GeV.} 
\label{beW} 
\end{center} 
\end{figure}

The shaded 
areas in the plots of Figures (\ref{beW}) and (\ref{abW}) correspond to regions that are excluded by the following  constraints: 
\begin{itemize} 
\item {\bf Vacuum stability:} 

Vacuum stability (at tree level) demands that 
 \begin{eqnarray} 
\lambda_{1,2}&>&0, \cr 
\lambda_{3},\lambda_{3}+ \lambda_{4}-|\lambda_{5}|&>& -2 
\sqrt{\lambda_{1}\lambda_{2}}. 
\label{vac_stab} 
\end{eqnarray} 
As a result, negative couplings, and $\lambda_L<0$  among others, are largely excluded. 
This is the 
shaded area in the domains $\mu_2 > M_{H_0}$ of Figures (\ref{beW}) and (\ref{abW}). 
\item {\bf Perturbativity:} 

Strong couplings $|\lambda_i|>4\pi$ are excluded but
intermediate couplings, $1~<~|\lambda_i|~<~4\pi$, might be tolerated. The latter correspond to the regions with horizontal lines. 
Notice that the mass splitting relative to the $\mu_2$ scale is 
smaller in the high mass regime (see Eq.(\ref{masses})). For this 
reason, going away from the diagonal line in the plots of Figure 
(\ref{abW}), we 
  run faster into the large coupling regime than in those of Figure (\ref{beW}). 
The regions excluded by the strong coupling constraint is of course symmetric 
 with respect to the $\mu_2 = M_{H_0}$ axis. 
\item {\bf Charged Higgs scalar:}  

The mass of the charged scalar $H^+$ is 
  constrained by LEP data to be larger than 79.3 GeV \cite{charged}. 
As we fix the mass differences in our plots, this constraint translates in  the 
excluded region $M_{H_0} \lsim 30$ GeV in the low mass regime abundance plots. 
\item {\bf Electroweak Precision Tests} (EWPT){\bf :} 

New physics can affect electroweak precision 
measurements. 
The impact of the new $H_2$ doublet can be described in term of the $S$, $T$
electroweak precision parameters. 
As pointed out in \cite{Barbieri:2006dq}, with appropriate mass splittings between its components, 
an $H_2$ could screen the contribution  to the $T$ parameter of a 
large Higgs masses, $M_h \sim 500$ GeV. 
The effects on the $S$ parameter is parametrically smaller for the region of parameter space 
satisfying the previous constraints. 
In all the case we have studied, we 
 have checked that the contribution to the $T$ parameter resulting from the Higgs mass and the $H_2$ components 
are consistent with EWPT. Unlike Barbieri {\em et al}, we only considered small values of the 
Higgs mass, $M_h \lsim 200$ GeV. 
Given our choice of the Higgs mass $M_h$ and of mass 
splittings between the $H_0$ particles and the other components of $H_2$, 
the constraints from EWPT are easily satisfied. 

\end{itemize} 
 
\noindent We now turn to the analysis of the model, starting with the low $H_0$ mass regime.

\subsection{Low mass regime, $M_H$ around 100 GeV} 
 
The results of this section are shown in Figure (\ref{beW}).  
Two processes  are relevant below the $W$, $Z$ or $h$ threshold: $H_0$ annihilation 
through the Higgs and $H_0$ coannihilation with $A_0$ through $Z$ exchange. 
Both give fermion-antifermion pairs, the former predominantly into $b\bar b$. 
As the mass of $H_0$ goes above $W$, $Z$ or $h$ threshold, $H_0$ annihilation
into  $WW$, $ZZ$ and $hh$ become 
increasingly efficient, an effect which strongly suppresses the $H_0$ relic density.\\ 
Coannihilation into a $Z$ may occur provided $\Delta MA_0$ is not
too important\footnote{$H_0\,H^+$ coannihilation is suppressed for our choice of $\Delta MH_c$.}, 
roughly $\Delta M$ must be of order of $T_{fo}\sim M_{H_0}/25$. 
Otherwise $H_0$ annihilate dominantly into 
$b \bar b$. 
This regime is quite interesting: the interactions are either completely 
 known ({\em i.e.} electroweak interactions), or highly sensitive to the Higgs boson mass $M_h$ 
and the $H_0 h$ effective coupling $\lambda_L$! 
 
For the sake of illustration, we study two 
cases, $M_h = 120$ GeV and  $M_h = 200$ GeV 
to show how the predictions change as a function of the Higgs boson mass. 
A few general lessons can be extracted from these two specific examples. 
Note that the mass differences $\Delta MA_0$ and  $\Delta MH_c$ are fixed respectively to 
10 and 50 GeV while $\Delta MA_0$ is large enough to avoid the constraints from direct detection and small enough 
   to have a some amount of 
   coannihilation into a $Z$. 
 
 \bigskip
\noindent
We may distinguish five different regions (see Figure (\ref{beW}), top left): 

\bigskip
Regions {\bf 1} and {\bf 2} are excluded respectively by charged Higgs production 
and vacuum stability constraints. The latter involves the self 
coupling of the Higgs, $\lambda_1$ and thus depends on the Higgs mass (see Eq.(\ref{hmass})): region {\bf 2} is broader 
  for smaller Higgs mass $M_h$. 
Although region {\bf 1} is experimentally ruled out, it is 
interesting to understand the trends. In this region, the relic 
abundance of $H_0$ decreases with increasing $M_{H_0}$ as one 
goes along $\mu_2 
  =M_{H_0}$. 
Comparing the $M_h= 120$ GeV and $M_h= 200$ GeV plots we observe that  the 
gradients depend, again, on the Higgs mass. 
This is simply due to the  $M_{H_0}^2/M_h^4$ dependence of the annihilation 
cross section through the Higgs, {\em i.e.} smaller abundance for 
higher $M_{H_0}$ and/or lower $M_h$. In region {\bf 2}, the 
abundance is smaller the further one deviates from $\mu_2 = 
M_{H_0}$, reflecting the dependence of the annihilation 
  cross section on $|\lambda_L|$. 
For $ M_{H_0} \lsim M_W$ (resp. $ M_{H_0} \gsim M_h$),
  $H_0 H_0\rightarrow \bar bb$ (resp. $H_0H_0\rightarrow h h$)
 is the dominant
 process. 
Otherwise, for $M_W\lsim M_{H_0}\lsim M_h$,  $H_0H_0\rightarrow W W,Z Z$ dominate.
 \bigskip

In region {\bf 3}, the couplings are rather large, but nevertheless 
consistent with vacuum stability. Since $\mu_2> M_{H_0},\, M_W$, the 
dominant contribution to the cross section is given by $H_0 
H_0\rightarrow W^+ W^-$ process, large enough to bring the relic 
abundance far below WMAP data. 
 
 \bigskip

Region {\bf 4} is below the $W$ threshold. It is the only region consistent with 
 the dark matter abundance predicted by WMAP data in the low mass regime. 
The process that determines the relic abundance of $H_0$ is again 
the annihilation through the Higgs. Coannihilation processes become 
dominant near the resonance at $M_{A_0}+M_{H_0} \approx M_Z$. This 
is the origin of the 
 dip in the $H_0$ relic abundance around $M_{H_0}\approx  40$ GeV (corresponding to $M_{A_0} \approx 50$ GeV
for our choice of mass splittings). 
Similarly, annihilation near the Higgs resonance generates a second 
dip 
 around $M_{H_0} \approx M_h/2$. 
There is an island consistent with WMAP data, which extends up to 
$M_{H_0}\sim 80$ GeV, at which point $WW$ annihilation
 becomes important and suppresses the relic abundance. 
 
 \bigskip

Region {\bf 5}, finally, corresponds to $M_{H_0}>M_W$. Annihilation into a gauge 
boson pair 
 is dominant, at least as  long as $M_{H_0}<M_h$. 
 Gauge boson pair production dominates above the Higgs threshold for $\mu_2 <M_{H_0}$ while Higgs pair production is dominant if 
 $\mu_2 >M_{H_0}$. In all instances, the relic abundance is too suppressed to be consistent with WMAP for 
$M_{H_0}\lsim 600$ GeV.

 \bigskip 
 
As expected, and as revealed by visual inspection, the photon flux 
plots shares some of the characteristics of the abundance plots. The 
only new salient feature is the absence of a dip at the Z resonance. 
This is of course because today, in contrast with the early 
universe, all $A_0$ are gone. Existing experiments are not very 
constraining but future gamma ray detection experiments, such as 
GLAST, might severely challenge the model. As usual we should bear in mind the astrophysical uncertainties: 
by changing the galactic dark matter density profile, we can get higher 
(or smaller) fluxes.

\bigskip 
The plot of the cross section for direct detection is pretty 
transparent. Suffices to notice (see section \ref{sec:DD}) that 
$\sigma_{H_0-p}$ at tree level is proportional to $\lambda_L^2$. The dominant
contribution to the elastic 
scattering cross section is thus zero on the $  \mu_2 =M_{H_0}$ axis while it 
increases for larger values of $\vert \mu_2 -M_{H_0}\vert$.

The dependence on the Higgs mass (Eq. \ref{eq:sigmaDMp})
clearly appears when comparing the $\sigma_{H_0-p}$ plots for $M_h= 120$ 
  GeV and  $M_h= 200$ GeV.
For $M_{H_0}>M_W$, the one-loop contribution to the cross section from $W$ exchange 
is taken into account. However it does not affect much
the results as it only amounts for $10^{-13}$ pb.
 Unless we 
suppose that the mass of $H_0$ and $A_0$ are nearly degenerate, 
existing direct detection experiments are not very constraining. 
Forthcoming experiments however might put a dent on some of the 
solutions consistent with WMAP.

 \subsection{The high mass regime, $M_H \gg M_W$} 

\begin{figure} 
\begin{center} 
\begin{tabular}{cc} 
\includegraphics[width=0.47\textwidth]{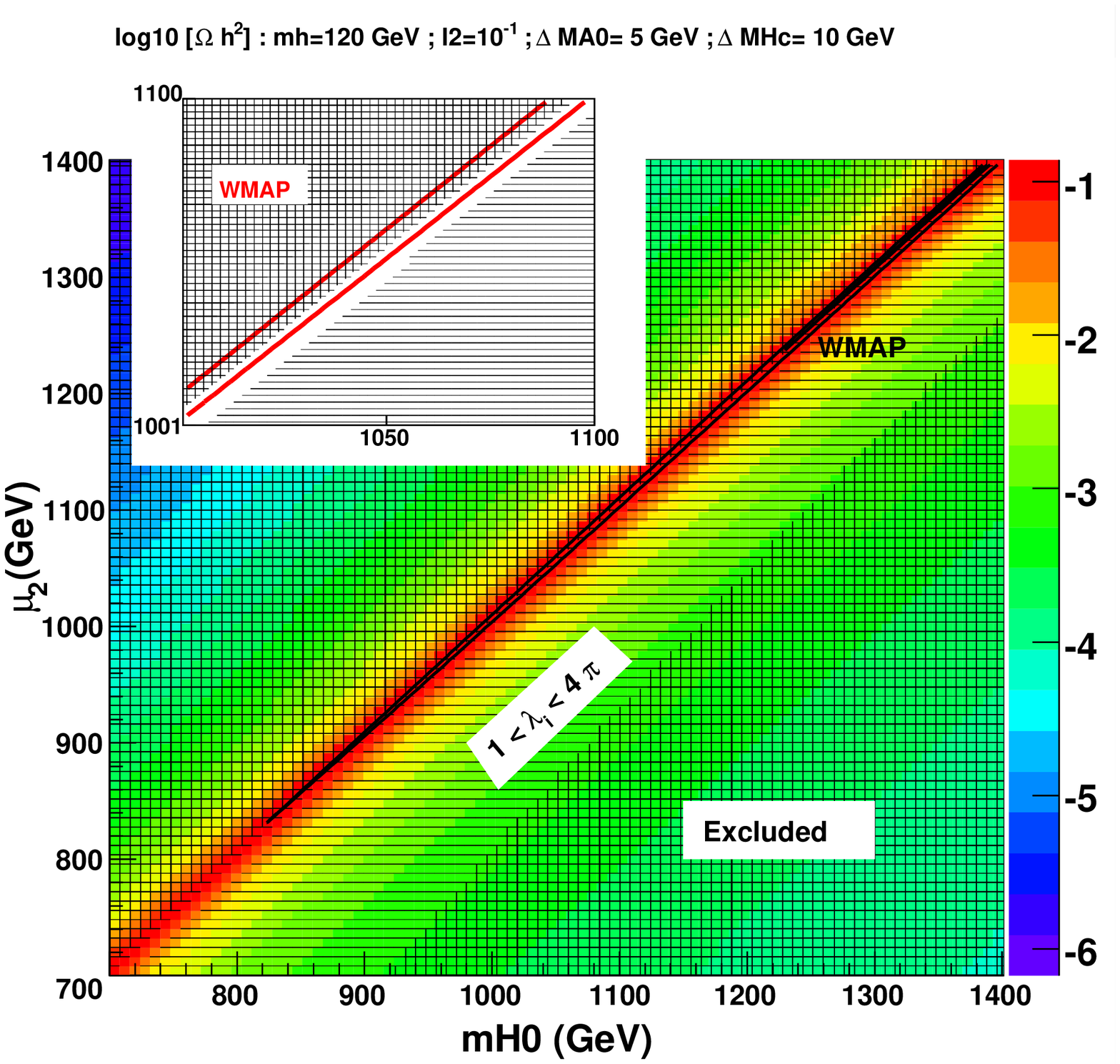}&\includegraphics[width=0.47\textwidth]{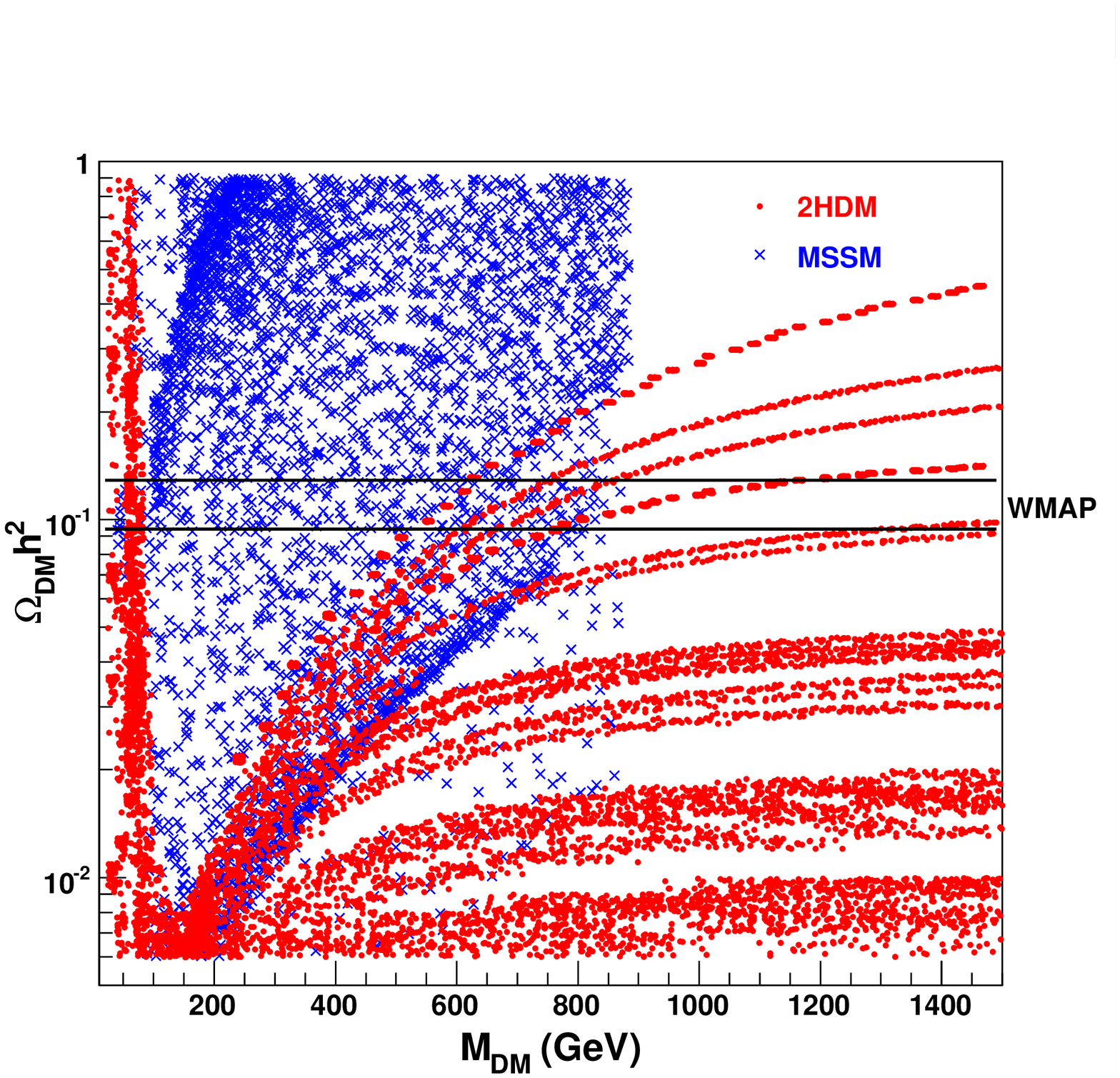}\\ 
\includegraphics[width=0.47\textwidth]{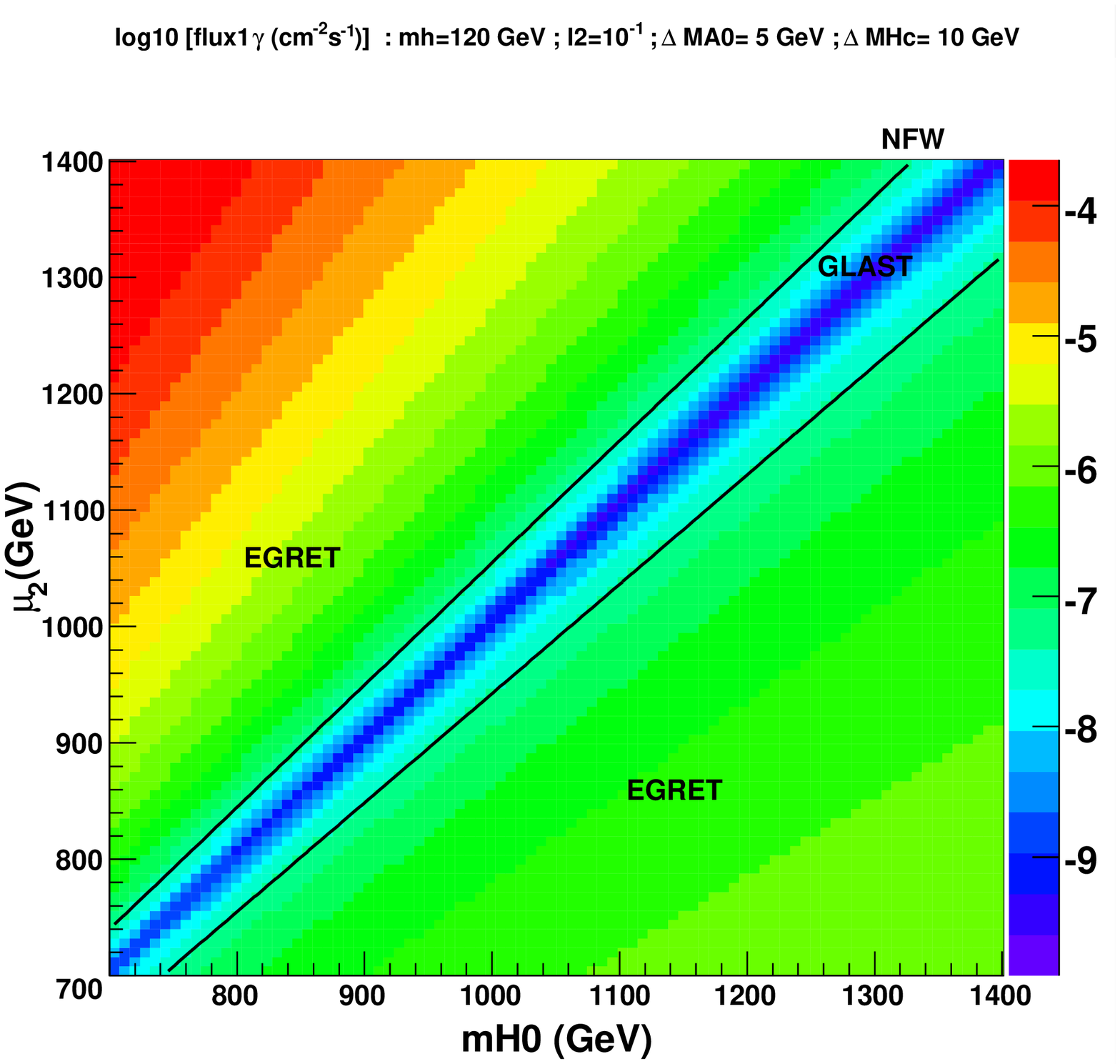}&\includegraphics[width=0.47\textwidth]{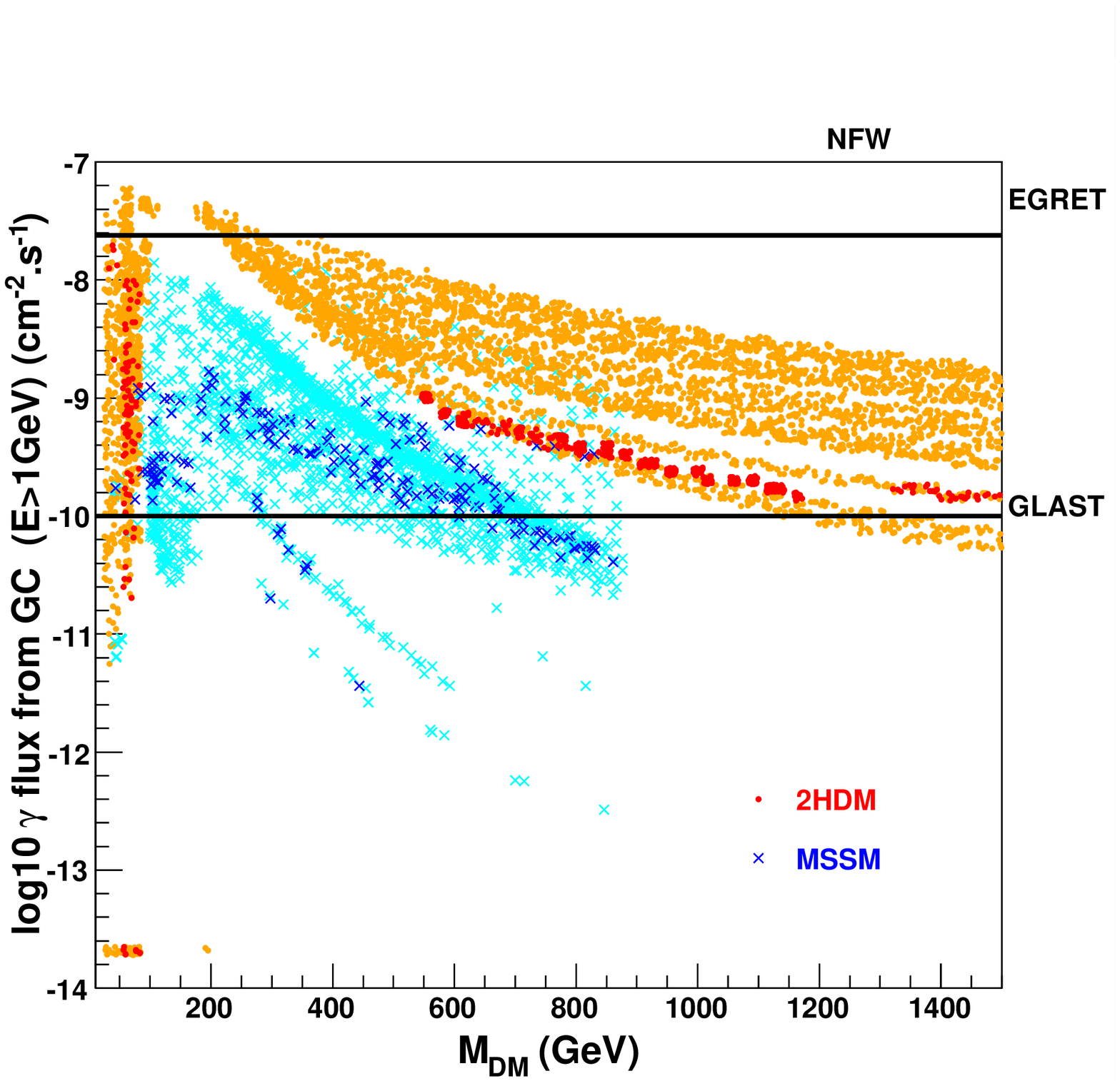}\\ 
\includegraphics[width=0.47\textwidth]{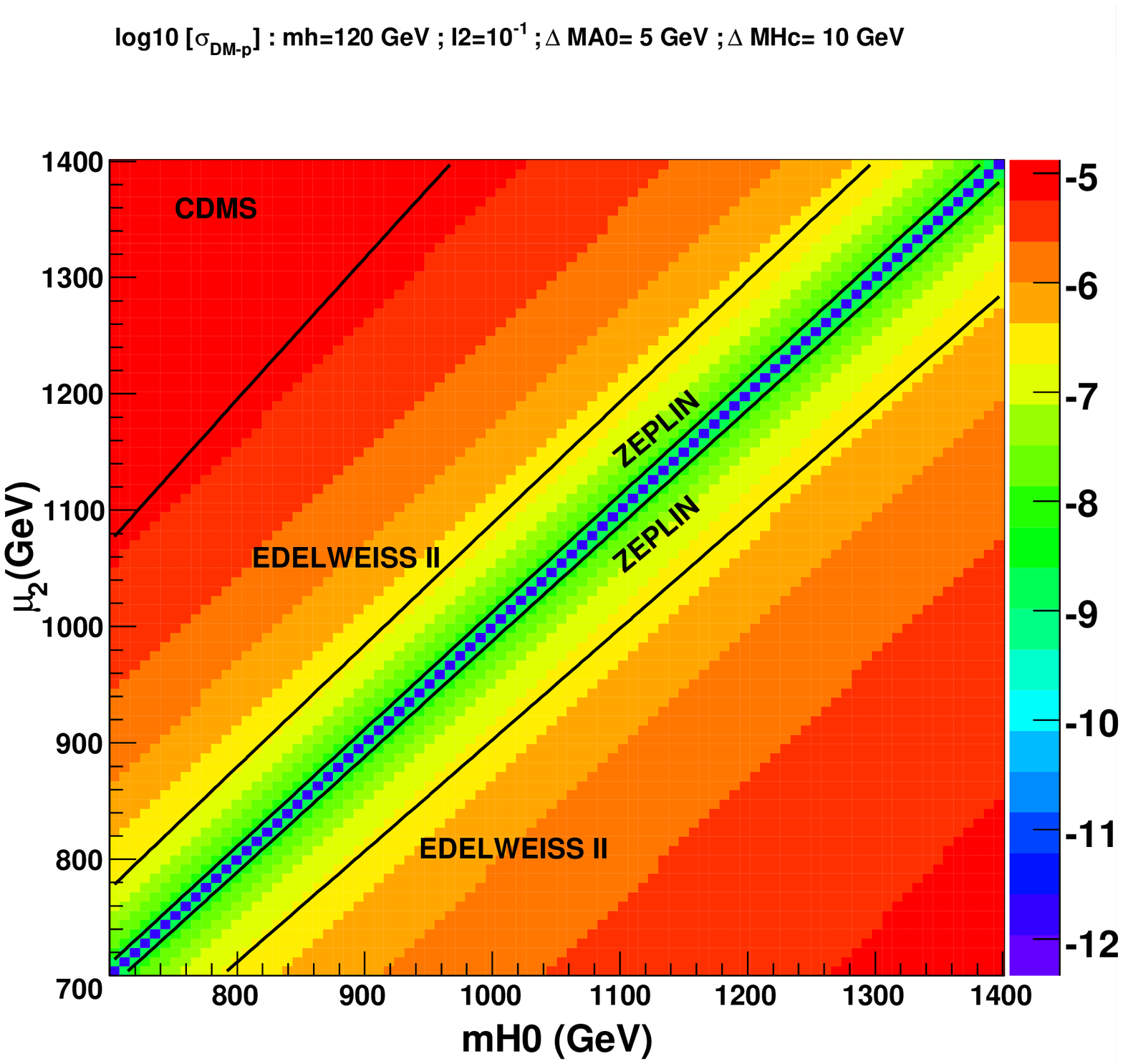}&\includegraphics[width=0.47\textwidth]{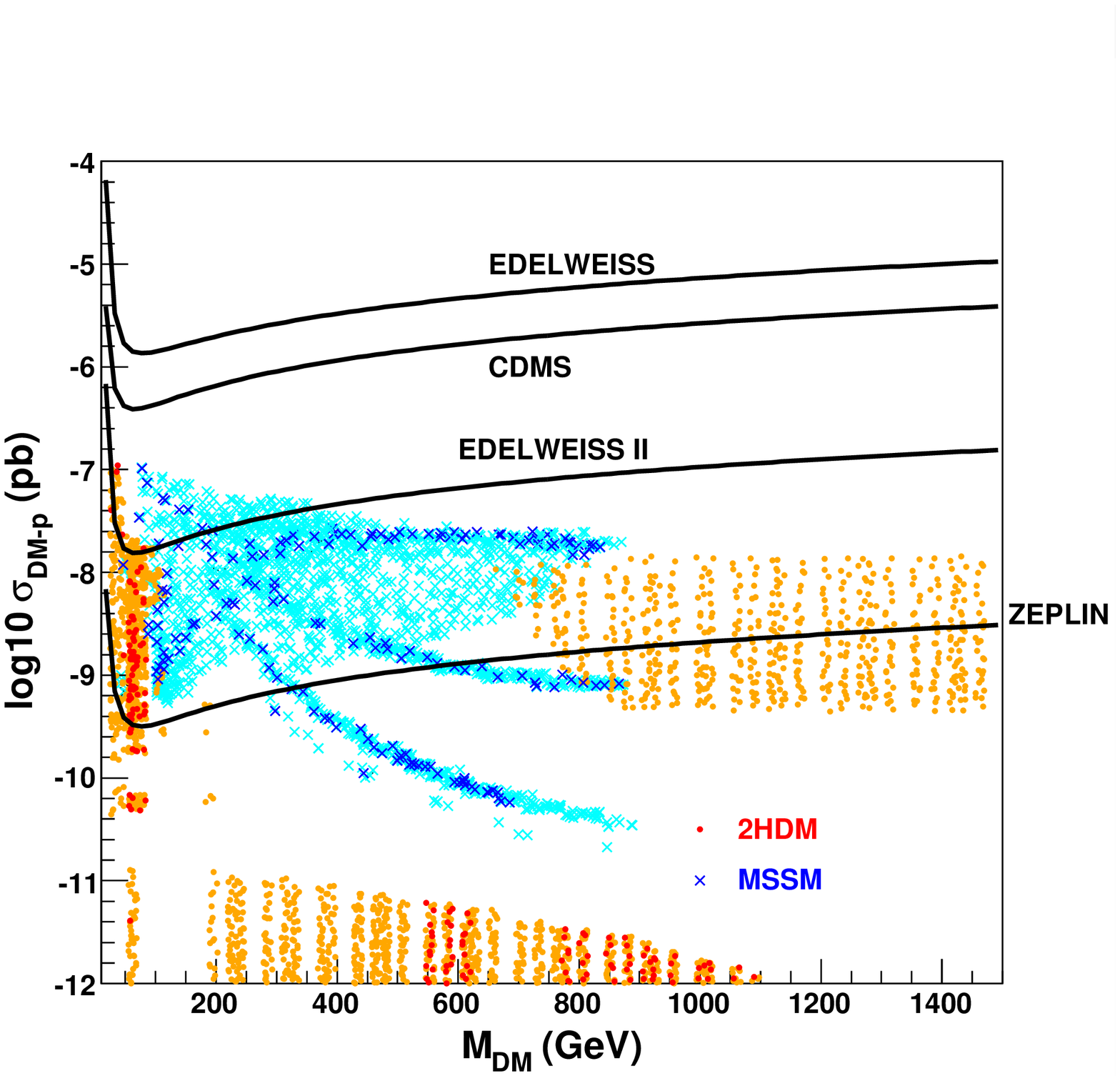} 
 \end{tabular} 
\caption{Left: same as Figure (\ref{beW}) for $M_{H_0},\mu_2 \in [700,1400]$
  GeV. Right: Comparison with the MSSM. From top to bottom: Relic density,
  gamma indirect detection, direct detection, all as a function of the mass of dark matter
  (for the last two plots, the light colors correspond to $0.01<\Omega_{DM}h^2<0.3$, while the dark
  colors correspond to $0.094<\Omega_{DM}h^2<0.129$.) In the direct detection plot, the two
 clouds of the IDM correspond to different values of 
$\lambda_L$, both in sign and amplitude.} 
\label{abW} 
\end{center} 
\end{figure}

The results of this section are shown in Figure (\ref{abW}), left column.
No new annihilation channel opens if $M_{H_0}$ is heavier than 
the Higgs or gauge bosons. There are then essentially two sort of 
processes which control both the abundance and the gamma ray 
flux: the annihilation into two gauge bosons, dominant if 
$\mu_2<M_{H_0}$, and the annihilation into two 
 Higgs, which dominates if $\mu_2>M_{H_0}$. 
 Coannihilation plays little role. It affects a bit the relic abundance along
 the diagonal
 but, even so, it is not the key process to get the WMAP abundance.

The abundance of dark matter is suppressed over most of the area of the
plot because of large quartic coupling effects on the cross sections. 
Strong couplings are excluded on a physical basis, but we found this limit nevertheless useful to understand the
interplay between the various processes. In the present subsection, we will argue
that it is possible to reach agreement with WMAP data, but only to the price of
some fine tuning between the different annihilation channels. 

\begin{figure}  
\begin{center}  
\begin{tabular}{cc}  
\includegraphics[width=0.47\textwidth]{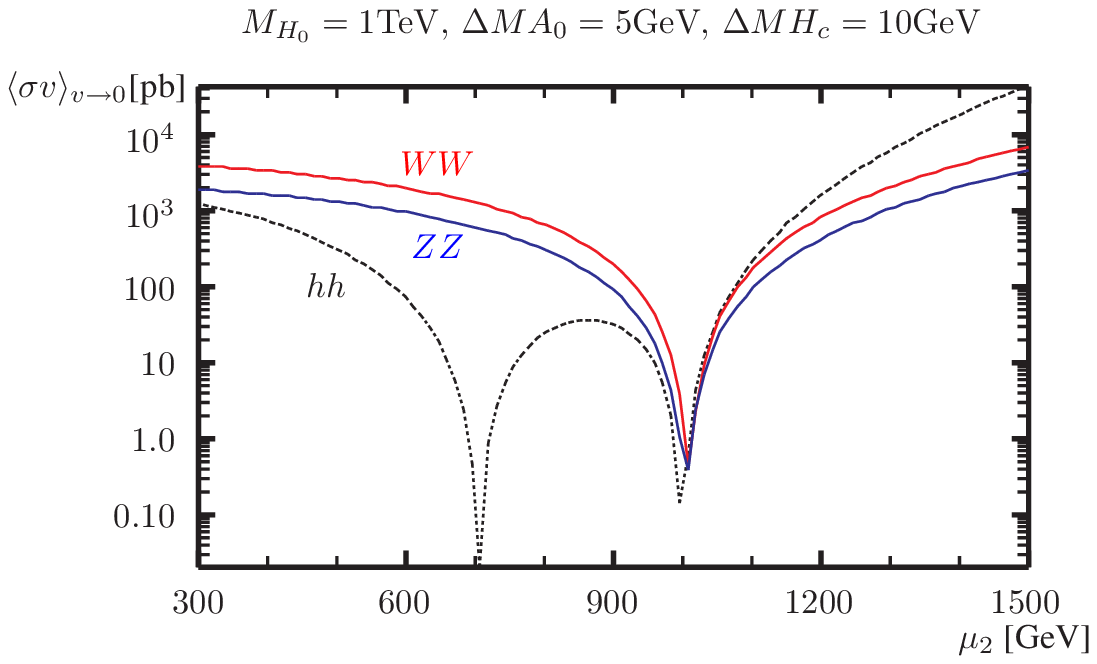}& 
\includegraphics[width=0.47\textwidth]{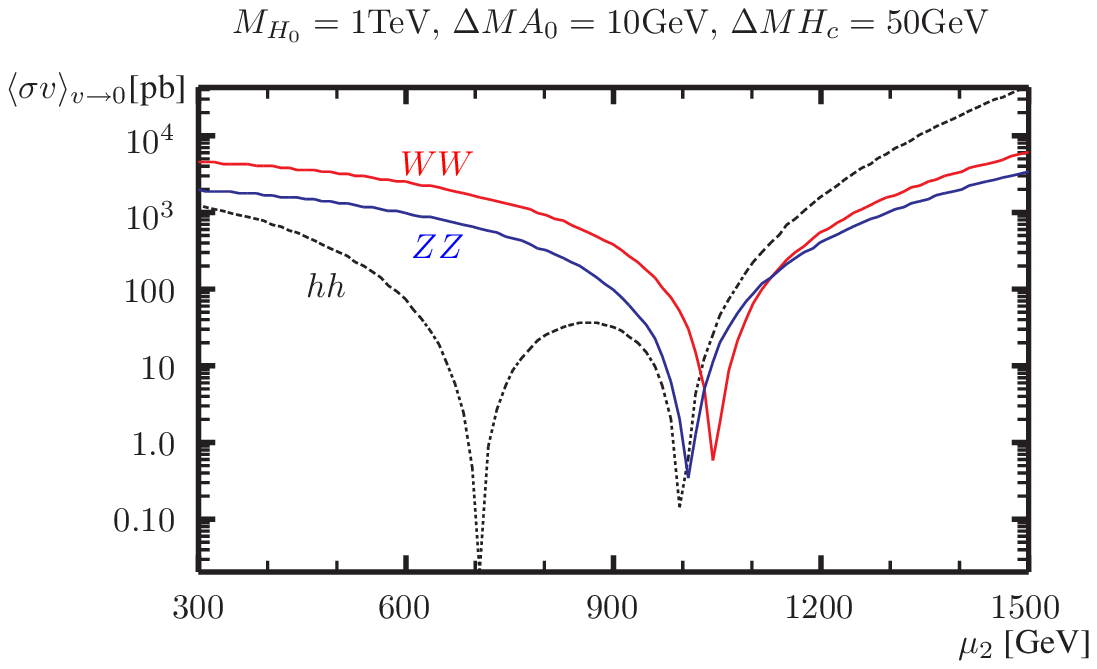}  
\end{tabular}   
\caption{$\langle\sigma v\rangle_{v\rightarrow 0}$ as a function of
$\mu_2$
  for $M_{H_0}=1000$  GeV and  small mass splittings 
$\Delta MA_0=5$ GeV, $\Delta MH+=10$ GeV for the plot on the left and
larger
  mass splittings  $\Delta MA_0=10$~GeV, $\Delta MH+=~50$~GeV  for the
 plot on the right. Dashed line corresponds to  
  annihilation into two Higgs, continuous lines to annihilation
into two  
  gauge bosons. }  
\label{fig:hhWWZZ}  
\end{center}  
\end{figure}

Consider first the annihilation into two Higgs bosons (the dashed line in the
graphs Figure (\ref{fig:hhWWZZ})).  Its cross section is vanishing for $\lambda_L = 0$. For
$\mu_2 < M_{H_0}$ (corresponding to $\lambda_L > 0$), there is a destructive interference
between the diagrams of Figure (\ref{feyn:higgs}), which is absent if $\mu_2 > M_{H_0}$.  
The annihilation into gauge bosons depends on the quartic
couplings between the scalars   (see Figure (\ref{feyn:WZ})). 
Indeed, the annihilation through an intermediate Higgs is
controlled by $\lambda_L$. 
Also, the t and u channels exchange
diagrams are sensitive to the mass differences
between the components of the $H_2$ doublet, which depend respectively on $\lambda_5$ for the annihilation
into 
$Z$ bosons and on $\lambda_4 + \lambda_5$ for the annihilation into $W$
bosons.

If $\lambda_L\simeq 0$, there is no or little annihilation into a Higgs. The
relevant diagrams are then
the quartic vertex with two gauge bosons and the t and u channels with $A_0$ (resp. $H^+$) exchange.
If $\lambda_5 = 0$ (resp. $\lambda_4 + \lambda_5 = 0$) the cross section into a $Z$ (resp. $W$) boson pair is minimal and
 scales like $\alpha_W^2/M_{H_0}^2$. 
If, for instance, $\lambda_5 \neq 0$ the annihilation into a $Z$ boson pair receives
a contribution which grows
like $\alpha_W^2 (M_{A_0} - M_{H_0})^2/M_Z^4$. 
A similar result holds for the annihilation into a $W$ boson pair provided $\lambda_4+\lambda_5\neq 0$.
Notice that  $\alpha_W^2 (M_{A_0} - M_{H_0})^2/M_Z^4 \sim \lambda_5^2/M_{H_0}^2$. 
The effects of weak isospin breaking between the components of $H_2$ is reminiscent of what happens for the Higgs 
in the regime of strong $\lambda_1$ coupling (the so-called Goldstone boson equivalence theorem \cite{Peskin:1995ev}). 

\bigskip

To have an abundance of dark matter in agreement with WMAP, the mass splittings between the components of
$H_2$ must be kept relatively small. First because large mass splittings correspond to large couplings and second because 
the different contributions to the annihilation cross section must be suppressed at the same location, 
around  $\lambda_L=0$ ({\it ie} $M_{H_0}\simeq \mu \simeq M_{A_0}\simeq
M_{H_+}$ in this case).
This is illustrated in Figure (\ref{fig:hhWWZZ}) for two different mass splittings.
The first plot is for small mass differences ($\Delta MA_0= 5$ GeV and $\Delta
MH_c= 10$ GeV), the second one for (relatively) larger splittings ($\Delta MA_0= 10$ GeV and $\Delta
MH_c= 50$ GeV).
In the second case, the cross section is too large to obtain an abundance consistent with WMAP,  $\langle\sigma
v\rangle_{\rm WMAP}\sim {\rm pb}$.

In the limit of small mass splittings and vanishing $\lambda_L=0$ we have the right amount of dark matter
provided $M_{H_0}\gsim 600$ GeV.
This regime corresponds to the narrow region around
the diagonal in Figure (\ref{abW}). The abundance increases for increasing $M_{H_0}$, but this can be somewhat 
compensated by playing with the
mass splittings, which, as discussed above, tend to increase the cross section. 
For instance, for $\Delta MA_0 = 5$ GeV and  $\Delta MH_c = 10$ GeV, we get the
right relic abundance for $M_{H_0}\gsim 800$ GeV.

There is however a limit to this trend.
Indeed, the total annihilation cross section 
of a scalar particle, like our $H_0$,  is constrained by unitarity to be 
smaller than 
\begin{equation} 
\label{unitarity} 
\langle\sigma
v\rangle_{v\rightarrow 0}^{\rm unit}\approx {4 \pi\over M_{H_0}^2}\sqrt{x_f\over \pi} 
\end{equation} 
a result which,  in the context of dark matter relics from 
freeze-out, 
 has been first put forward by Griest and Kamionkowski \cite{Griest:1989wd}. 
In the present model, increasing the mass splitting drives the annihilation cross section 
toward the strong quartic coupling regime. A similar result 
 holds, for instance, for a weakly interacting massive neutrino candidate of dark matter. In this case, both the neutrino and 
its charged partner should be kept lighter 
 than, say $1$ TeV, since their mass comes from Yukawa
 couplings to the Higgs field. (See \cite{Enqvist:1988we} and the discussion in \cite{Griest:1989wd}). In our case, the bulk of the mass
 of the components of the $H_2$ doublet comes from the mass scale $\mu_2$, which is {\em a priori} arbitrary. However,  
the unitarity limit and WMAP data constraints, 
translate into the
 upper bound $M_{H_0}\lsim  130$ TeV \cite{Griest:1989wd}.

\bigskip 
 
Unfortunately neither direct, nor indirect detection experiments are sensitive 
to the large mass region discussed in this section. Forthcoming 
direct detection experiments might do better, but as 
the plots show rather clearly, other forms of dark matter would then have 
to be introduced in order to explain the amount of dark matter that is currently observed. However, if the profile of dark matter in the galaxy is as assumed
in the present paper ({\em i.e.} NFW), future gamma detectors (GLAST) will probe most of 
the parameter space considered in this section, keeping in mind that we had to finely 
tune the mass splittings to obtain the right abundance.

\section{Conclusions}\label{sec:conclusions} 
 
The dark matter candidate of the Inert Doublet Model stands fiercely 
by the neutralino. 
The lightest stable scalar is   
a weakly interacting massive
particle with a rich, yet simple, phenomenology and it has 
a true potential for being constrained by existing and forthcoming experiments looking for 
dark matter.  For the sake of comparison, 
 we show in the right part of Figure (\ref{abW}) a
 fair sample of models, both for the IDM and the MSSM.  
In the $(M_{DM},\Omega_{DM} h^2)$ plane, we clearly see the two regimes (low mass and high mass) of the IDM that may give
 rise to a relevant relic density
({\it i.e} near WMAP).  The MSSM models have 
a more continuous behavior, with 
${\cal O}(100$ GeV) dark matter masses. 
For indirect detection, the IDM dark matter candidates have typically higher detection rates than the neutralino in 
SUSY models, especially at high mass. It is in particular interesting that the IDM  
can give the right relic abundance in a range of parameters which will be probed by GLAST. Direct detection 
is also promising but only at low masses, 
where the models are within reach of future ton-size experiments. The
high mass models, however, have too small cross sections in the regions consistent with WMAP. 

The phenomenology of the IDM candidate for dark matter is intertwined with that of the 
Brout-Englert-Higgs particle, $h$. It could be of 
some 
interest to investigate the prospect for detection of the $h$ and $H_2$ components at the LHC,
in the light of the constraints
for dark matter discussed in the present paper. 
   
   \section*{Acknowledgments}

This work is supported by the FNRS, the I.I.S.N. and the
Belgian Federal Science Policy (IAP 5/27).

\end{document}